\input harvmac

{\Title{\vbox{
\hbox{%
}
}}
{\vbox{
\centerline{BPS Wall Crossing and Topological Strings }}}
\vskip .3in
\centerline{Sergio Cecotti$^1$ and Cumrun
Vafa$^{2}$}
\vskip .4in
}

\centerline{$^1$ Scuola Internazionale Superiore di Studi Avanzati via Bierut 2-4 I-34100 Trieste, Italy}
\centerline{$^2$ Jefferson Physical Laboratory, Harvard University,
Cambridge, MA 02138, USA}

\vskip .4in

By embedding $N=2$ gauge theories in string theory
and utilizing string dualities we map the counting of BPS
states with arbitrary electric and magnetic charges to computations of an A-model topological string
on an associated geometry constructed from the data of the SW curve.
We show how the conjecture of Kontsevich and Soibelman
regarding wall crossing,
as well as a more refined version which captures the spin content of BPS states,
is a natural consequence.
Chern-Simons theory realized on A-branes and a twistorial construction play key roles.

\vfill
\eject

\newsec{Introduction}

The study of BPS states has been a key ingredient in understanding
quantum aspects of supersymmetric theories.  In particular one can
obtain exact information about their masses and degeneracies as has been
known for a long time \ref\WittenMH{
  E.~Witten and D.~I.~Olive,
  ``Supersymmetry Algebras That Include Topological Charges,''
  Phys.\ Lett.\  B {\bf 78}, 97 (1978).
  %%CITATION = PHLTA,B78,97;%%
}.  It is thus
not surprising that BPS objects have played a key role in developing string
dualities.

One of the important facts about BPS states, which was discovered
also a long time ago
\ref\cecva{
  S.~Cecotti and C.~Vafa,
  ``On classification of N=2 supersymmetric theories,''
  Commun.\ Math.\ Phys.\  {\bf 158}, 569 (1993)
  [arXiv:hep-th/9211097].
  %%CITATION = CMPHA,158,569;%%
},
 in the context of supersymmetric $(2,2)$ theories
in 2 dimensions, is that their degeneracy can
jump when the central charges are given by points on the complex plane.
When the phases of two BPS particles interchange, then they can disappear
(or new ones may appear).
 At the face of it this sounds contradictory with continuity
of correlation functions in path-integrals.  However
it was shown in \ref\cfiv{
  S.~Cecotti, P.~Fendley, K.~A.~Intriligator and C.~Vafa,
  ``A New supersymmetric index,''
  Nucl.\ Phys.\  B {\bf 386}, 405 (1992)
  [arXiv:hep-th/9204102].
  %%CITATION = NUPHA,B386,405;%%
}\ how this contradiction
is resolved:  When a BPS particle decays after crossing a wall
what used to be a single BPS particle contribution to correlation
functions gets replaced by multi-particle contributions. Moreover
the continuity of a specific path-integral computation, captured by the
index
$${\rm Tr}(-1)^{\rm F} {\rm F} \ {\rm exp}(-\beta H)$$
can be used to give a general recipe as to how the
degeneracy of BPS particles jump as we cross the wall.

In the context of ${ N}=2$ supersymmetric theories in four
dimensions, the central charges
of electric and magnetic particles are also given by points
on the complex plane, and thus there can be jumps.  In
fact it was observed by Seiberg and Witten \ref\SeibergRS{
  N.~Seiberg and E.~Witten,
  ``Monopole Condensation, And Confinement In N=2 Supersymmetric Yang-Mills
  Theory,''
  Nucl.\ Phys.\  B {\bf 426}, 19 (1994)
  [Erratum-ibid.\  B {\bf 430}, 485 (1994)]
  [arXiv:hep-th/9407087].
  %%CITATION = NUPHA,B426,19;%%
}\ that not only there can be jumps, but that there has to be
jumps for the theory to be consistent.  In particular they
obtained a very intricate pattern of BPS jumps which are far
more complicated than the ones encountered in the 2d theories.
Namely in such cases sometimes infinitely many BPS particles disappear
as one crosses the wall.  That this could happen follows because
the BPS central charges lie on the projection of an integral lattice
to the complex plane and thus when any two get aligned, infinitely
many central charges get aligned due to linearity.
However there was no systematic way to compute these degeneracies
and their jumps.

In the context of realizing $N=2$ gauge theories in 4 dimensions, with
the advent of string dualities,
the situation became a bit more clear:  It was realized in
\ref\klmvw{
  A.~Klemm, W.~Lerche, P.~Mayr, C.~Vafa and N.~P.~Warner,
  ``Self-Dual Strings and N=2 Supersymmetric Field Theory,''
  Nucl.\ Phys.\  B {\bf 477}, 746 (1996)
  [arXiv:hep-th/9604034].
  %%CITATION = NUPHA,B477,746;%%
}\ that by geometric engineering
and an application of a series of dualities, $N=2$ gauge systems are mapped
in the type IIA setup to NS 5-branes whose worldvolume is
${\bf R}^4\times \Sigma$ where $\Sigma$ is the Seiberg-Witten curve.
Moreover the BPS states were mapped to D2 branes ending on the
5-brane.  Using this, it became possible to compute
the BPS degeneracies directly, at least in some simple
situations, and see that indeed they jump according to
what was expected \klmvw \ref\shapv{
  A.~D.~Shapere and C.~Vafa,
  ``BPS structure of Argyres-Douglas superconformal theories,''
  arXiv:hep-th/9910182 .}.  For certain situations, a wall-crossing
formula was proposed in 
\ref\DenefVG{
  F.~Denef and G.~W.~Moore,
  ``Split states, entropy enigmas, holes and halos,''
  arXiv:hep-th/0702146.
  %%CITATION = HEP-TH/0702146;%%
}. 
 However it
was not possible to extend this to the general case.

Recently a surprising conjecture was put forward by
Kontsevich and Soibelman \ref\KS{M.~Kontsevich and Y.~Soibelman,
``Stability Structures, Motivic Donaldson-Thomas Invariants and Cluster Transformations,"
[arXiv:0811.2435].}\ as to how 
the BPS degeneracies jump.  It was shown in \ref\GMN{
  D.~Gaiotto, G.~W.~Moore and A.~Neitzke,
  ``Four-dimensional wall-crossing via three-dimensional field theory,''
  arXiv:0807.4723 [hep-th]\semi
  D.~Gaiotto, G.~W.~Moore and A.~Neitzke,
  ``Wall-crossing, Hitchin Systems, and the WKB Approximation,''
  arXiv:0907.3987 [hep-th].}
how this result can be derived by demanding that the
contribution of BPS particles to hyperk\"ahler metrics
obtained upon compactification on a circle, be
continuous.  Again, as in the 2d case, the continuity
of the correlation functions was due to multi-particle contributions
replacing single particle contributions.

In this paper we offer an alternative derivation of
the KS conjecture for all the $N=2$ theories that can be associated
to a Seiberg-Witten curve.\foot{There are some solvable $N=2$
systems, such as
quiver theories based on E-groups
\ref\kmv{
  S.~Katz, P.~Mayr and C.~Vafa,
  ``Mirror symmetry and exact solution of 4D N = 2 gauge theories. I,''
  Adv.\ Theor.\ Math.\ Phys.\  {\bf 1}, 53 (1998)
  [arXiv:hep-th/9706110].}\ which suggest that the
corresponding $N=2$ geometry is not given in general by a Seiberg-Witten
curve, but rather by a local Calabi-Yau threefold.}  In fact we go further:
We show that the computation of BPS degeneracies can be mapped
to a computation
of open topological A-model strings, for which a great deal is known.
More precisely, we compactify this IIA theory
on a circle and use an 11/9 flip which
changes which direction corresponds to the circle of M-theory.
In other words we lift the 4d type IIA theory to M-theory,
in which case the NS 5-brane gets mapped to M5 brane, and compactify
the theory on a circle, which we now view as the 11-th circle.
In this way we get back to IIA theory, but now the NS 5 brane
is replaced by a D4 brane whose worldvolume is ${\bf R}^3\times \Sigma$.
In this context the BPS particles correspond to $D2$ branes ending
on $D4$ branes.  This is close to the situations
where we know how to count
their degeneracy \ref\oogv{
  H.~Ooguri and C.~Vafa,
  ``Knot invariants and topological strings,''
  Nucl.\ Phys.\  B {\bf 577}, 419 (2000)
  [arXiv:hep-th/9912123].}:  They correspond
to partition functions of topological A-model strings where
D4 brane wraps over a Lagrangian cycle of Calabi-Yau threefold.
There is one difference however:  We are in a situation corresponding
to twice as much supersymmetry.  To remedy this, motivated by the twistor construction,
we compactify on a
circle down to 2 dimensions, where
the Seiberg-Witten curve fibers in an interesting way
over the circle.
In this way we can break half of its supersymmetry.  We are then
down to 2 dimensions, where D4 brane wraps
$\Sigma \times S^1 \times {\bf R}^2$ and topological string computes specific
F-term corrections to the $(2,2)$ supersymmetric theory in 2 dimensions
as in \oogv.  Moreover these get mapped to computation of correlation
functions of a $U(1)$ Chern-Simons theory living on $\Sigma \times S^1$.
The BPS particles ending on the D4 brane induce Wilson Loop operators
on the Chern-Simons theory.  Moreover
we choose the fibration structure of $\Sigma$ over $S^1$ such
that each BPS particle with a given phase for the central
charge attaches to the D4 brane at the corresponding
angle on the circle!  Viewing the circle
as the time direction, leads to the computation of the path-integral
of Chern-Simons theory by taking the trace of these operator in a
time ordered fashion; and since time and phase are identified here,
it means ordering in terms of the phase of their BPS charge.  The fact
that partition function of the A-model topological
strings can change only in a continuous way as we
change parameters, leads directly to the results of KS \foot{
The idea to use topological strings as a tool for gaining
insight into wall crossing
phenomena has been considered in a different context in
\ref\CollinucciNV{
  A.~Collinucci, P.~Soler and A.~M.~Uranga,
  ``Non-perturbative effects and wall-crossing from topological strings,''
  arXiv:0904.1133 [hep-th].
  %%CITATION = ARXIV:0904.1133;%%
}.}.  One way
to phrase our results is the following:  The BPS degeneracies
lead to a particular (infinite) combination of link invariants on $\Sigma
\times S^1$.  As we
cross the walls of marginal stability the links change topological type
and thus the associated Chern-Simons invariants change.  However, the BPS degeneracies
which dictate which combinations of the invariants we take also
change in such a way that the full amplitude does not change!

We get two further refinements:  First of all, it
is known that the spin of the BPS particles are also encoded
in the topological string in terms of the coupling constant of topological
string.  Using this we get a refinement of KS conjecture, which
was already suggested in \ref\guktudor{
  T.~Dimofte and S.~Gukov,
  ``Refined, Motivic, and Quantum,''
  arXiv:0904.1420 [hep-th]
\semi
  T.~Dimofte, S.~Gukov and Y. Soibelman, to appear.}.  Moreover
we get a further refinement, as in \oogv \ref\marv{
  M.~Marino and C.~Vafa,
  ``Framed knots at large N,''
  arXiv:hep-th/0108064.
  %%CITATION = HEP-TH/0108064;%%
}\
by considering more than one D4 brane.
Namely there is a further refined
integrality which is encoded
in topological string partition function having to do with the fact that a
D2
brane can end on multitude of boundary circles on $\Sigma$,
in terms of Wilson loop correlations of a $U(N)$ Chern-Simons
theory.

The organization of this paper is as follows:  In section 2 we review
aspects of topological A-model string.  In particular we discuss
integrality properties of open and closed string amplitudes
as well as the relation to Chern-Simons theory and Wilson loop
operators.  In section 3 we discuss 5-brane geometries which
realize $N=2$ systems in 4d as well as their further compactifications.
We discuss a general class of twistorial compactifications which
is relevant for our application.  In section 4 we use this to
derive the KS conjecture and its refinement. In section 5 we 
attempt to explicitly construct such branes, where we 
point out that  the most obvious construction,
even though does not lead to Lagrangian branes,
is sufficiently close to one, to establish our result.   Finally in section 6 we
end with some concluding thoughts.

\newsec{Aspects of A-model Topological Strings}
Topological A/B strings compute F-terms for type IIA/B compactifications
on Calabi-Yau threefold.  In this paper we would be mainly interested
in type IIA case, for reasons that will become clear.  Here we
review some aspects of A-model topological strings that we need.

A-model topological string \ref\WittenIG{
  E.~Witten,
  ``On The Structure of the Topological Phase of Two-Dimensional Gravity,''
  Nucl.\ Phys.\  B {\bf 340}, 281 (1990).}\ `counts'
holomorphic maps from string worldsheet to Calabi-Yau threefolds.
It is known that due to the fact that generically these maps come with moduli, the counting in general would lead to rational numbers.  However
it was shown in \ref\govai{
  R.~Gopakumar and C.~Vafa,
  ``M-theory and topological strings. I,''
  arXiv:hep-th/9809187 \semi
  R.~Gopakumar and C.~Vafa,
  ``M-theory and topological strings. II,''
  arXiv:hep-th/9812127.}\ that there is an integrality
in this expansion given by
$$ Z={\rm exp}\big[\sum_{Q,j} N_{Q,j}\sum_n {1\over n}
e^{-nt\cdot Q}
(q^{n/2}-q^{-n/2})^{2j-2}\big]$$
where $q={\rm exp}(-g_s)$, $t$ is the Kahler moduli and $N_{Q,j}$ are integers that
count the number of BPS branes in one higher dimension.  More precisely,
consider M-theory on the same Calabi-Yau in one higher dimension.  Then $N_{Q,j}$ counts the
net numeber of BPS $M2$ branes in the class $Q\in H_2(CY,{\bf Z})$ and
$SU(2)_L$ spin $j$, where $SU(2)_L\times SU(2)_R $ is the rotation
group in 5 dimensions. This derivation used the duality of IIA theory with M-theory, where the M2 branes contribute to F-terms of type IIA theory on Calabi-Yau
generated by BPS particles going around the M-theory circle.

One can also consider open topological strings.  In that case
one introduces D-branes wrapping Lagrangian submanifolds
\ref\wittencs{
  E.~Witten, ``Chern-Simons Gauge Theory As A String Theory,''
  Prog.\ Math.\  {\bf 133}, 637 (1995)
  [arXiv:hep-th/9207094].}\ and considers holomorphic
maps from the worldsheet with boundary, where the boundary
lies on the D-brane.  Moreover it was shown in \wittencs\
that on the topological D-brane lives a Chern-Simons theory.
Moreover for every holomorphic curve of area $A$ ending on a loop $\gamma$ in the
D-brane, one gets an insertion of the Wilson loop in the Chern-Simons
Lagrangian, which schematically looks as
$U_{\gamma} {\rm exp}(-A) $, where $U_{\gamma}$
is the holonomy of the gauge field around $\gamma$.  There are also integrality
properties for these corrections to Chern-Simons theory which
have a similar structure to that of the closed string sector \oogv .
For simplicity let us consider the case where we have one D-brane
wrapping a Lagrangian cycle with one non-trivial 1 cycle.  The generalization
to when we have more than 1-cycle is straight-forward.  Also let us assume that
there are no non-trivial 2-cycles in the CY (as will
be the main focus of this paper).  The general case is discussed in \oogv . Let $U$ denote
the holonomy of the $U(1)$ gauge field around that cycle.  Then
the open topological string partition function is given by
\eqn\opbps{Z={\rm exp}\Big[ \sum_{j,m} N_{j,m} \sum_n{U^{nm}\over n} {q^{nj}\over (q^{n/2}-q^{-n/2})}\Big] }
where $N_{j,m}$ are some integers counting certain BPS states in the
dual M-theory.  
More precisely, consider type IIA string with D4 branes wraping the
Lagrangian cycle of CY and filling a 2d subspace of spacetime.  This lifts
in M-theory to an M5 brane wrapping the Lagrangian cycle and filling
${\bf R}^3$.  $N_{j,m}$ counts the net number of M2 branes ending on
$M5$ brane, whose boundary wraps the non-trivial cycle 
of the Lagrangian brane $m$ times and has spin $j$ in ${\bf R}^3$.
Note that if we have more than one-cycle in the Lagrangian
submanifold we get similar contributions for each cycle and
for each M2 brane ending on the M5 brane.
Furthermore this counting can be refined if we consider $K$ instead
of 1 brane wrapping the Lagrangian brane.  This integrality
can also be expressed in similar terms \oogv \marv .
For more than one Lagrangian brane, the label $m$
gets replaced by the representation lablel ${\cal R}$ and the $(U^m)^n$ get replaced
by ${\rm tr}_{\cal R}U^n$.

If we have more than one cycle in the Lagrangian submanifold, as will be the case for
applications we will be interested, we get operator insertions in the CS theory for M2 branes,
ending in a class $\gamma$ , and of spin $j$ given by
\eqn\useful{O_{j,\gamma}={\rm exp}\Big[N_{j,\gamma}\sum_n{U_{\gamma}^{n}\over n} {q^{nj}\over (q^{n/2}-q^{-n/2})}\Big]}
Note that these are insertions in the CS path integral, and so we need to compute the partition
functions of the CS theory with these insertions to get the full amplitude for A-model topological
strings in the presence of Lagrangian A-branes.

\newsec{Engineering $N=2$ Theories in 4d and Their Compactifications}

In this section we consider the chain of dualities
which map the 4d $N=2$ theory to an associated 2d theory with $(2,2)$
supersymmetry.  In the next section we will use this
setup to show how the BPS degeneracies can be recovered from open
A-model topological strings.  As a byproduct, this leads to a simple derivation of a refined version of KS conjecture.

We will start with the theory in 4d realized by an M5 brane
wrapping $\Sigma \times {\bf R}^4$ where $\Sigma$ is a curve defined
by \foot{ Note that
these classes of 5-branes span a large class of known $ N=2$
theories:  In particular it is known that all the $N=2$ theories
geometrically engineered in type IIA using 3-folds are of this type \kmv , where by mirror symmetry we get a type IIB theory on the geometry
$$uv+F(z_1,z_2)=0,$$
which in turn using T-duality \ref\OoguriWJ{
  H.~Ooguri and C.~Vafa,
  ``Two-Dimensional Black Hole and Singularities of CY Manifolds,''
  Nucl.\ Phys.\  B {\bf 463}, 55 (1996)
  [arXiv:hep-th/9511164].
  %%CITATION = NUPHA,B463,55;%%
}, gets mapped
to NS 5-brane
on $F(z_1,z_2)=0$ \klmvw .  This can be lifted to
M-theory where the NS5 brane becomes the M5-brane wrapping this curve.
Alternatively we can also consider brane constructions
of $N=2$ theories as in \ref\WittenSC{
  E.~Witten,
  ``Solutions of four-dimensional field theories via M-theory,''
  Nucl.\ Phys.\  B {\bf 500}, 3 (1997)
  [arXiv:hep-th/9703166].
}\ which leads directly to such geometries for constructing
$N$=2 theories (see also the recent work \ref\Gai{
  D.~Gaiotto,
  ``N=2 dualities,''
  arXiv:0904.2715 [hep-th].
}).}
$$F(z_1,z_2)=0, \qquad  (x,z_3)=0.$$
where $x$ is a real line, $z_3$ is a complex plane, and
for simplicity we assume $(z_1,z_2)\in {\bf C}^2$.

In this setup the BPS states correspond to M2 branes ending on
$\Sigma$.  Let $D$ denote such an M2 brane with  $\partial D \subset \Sigma$.  The central charge of the ${ N}=2$ theory for the M2 brane is given by
$$Z=\int_D i\,dz_1\wedge dz_2$$
and its mass is given by
$$|Z|=\int_D i\,e^{-i\theta}\, dz_1\wedge dz_2$$
In particular $e^{i\theta}$ measures the phase of $i\,dz_1\wedge dz_2$ restricted
to the M2 brane.
Note in particular that fixing the $\theta$ for which there is
a BPS particle picks out a given ray in $H_1(\Sigma)$, which
corresponds to that given by $\partial D\in H_1$.  
This implies that on the boundary $\partial D$ the one form $\lambda =z_2dz_1$, which is
identified with the Seiberg-Witten differential has a definite phase.  Viewing $z_2(z_1)=W'(z_1)$  maps this boundary curve
on $\Sigma$ to a straight line in the $W$ plane, just as is the case
 in the $N=2$ LG theory in 2d \cecva .  This fact was used
to capture BPS degeneracies in some simple cases \klmvw \shapv ,
and confirm some of the expected BPS jumping phenomena
as one crosses the walls of marginal stability.  However, 
in order to find a more systematic way to capture the BPS
degeneracies we need some new ideas, which we will
now turn to.

Consider compactifying this theory on a circle, down
to 3 dimensions.  However, now we view the 4th dimension
as the M-theory direction.  This leads to a type IIA theory
where we have a D4 brane wrapping $\Sigma$ and filling
the ${\bf R}^3$ spacetime.  Next we consider further
compactification on a circle:
$${\bf R}^3\rightarrow S^1\times {\bf R}^2.$$
We thus end up with a theory living on the D4 brane in 2 dimensions with
$(4,4)$ supersymmetry.  To see the structure of this theory better, let us combine
this $S^1$ with the $x$-line, and call it a new
cylindrical coordinate $\zeta ={\rm exp}(i\theta -x)$.
We can view this new geometry as a non-compact flat Calabi-Yau, given
by
$$(z_1,z_2,\zeta )$$
with the rest of the spacetime being
$$(z_3,{\bf R}^2 )$$
where D4 brane wraps the ${\bf R^2}$ (at $z_3=0$) and a 3d internal
submanifold
$$F(z_1,z_2)=0,\qquad |\zeta|=1.$$
Since ${\bf C}^2$ is hyperk\"ahler, by a rotation
of complex structure we can view this 3d submanifold as
a Lagrangian submanifold of the CY.  On the non-compact
worldvolume of D4 brane
we get a $(4,4)$ supersymmetric theory in 2 dimensions.

We now consider the topological A-model in the internal
space.  This would have worldsheet instantons which wrap
holomorphic curves in ${\bf C}^2$ which end on $\Sigma \times S^1$.
All such curves are at a fixed value of $\theta \in S^1$.
These are precisely the BPS particles we are after!
More precisely, as in \oogv\ we know that the degeneracy
and the spin of M2 branes ending on M5 branes are captured
by topological strings, as reviewed in the last section.
However, the story is not so simple:  Here we have
twice as much supersymmetry compared to the one discussed
in \oogv\ because the space is flat (or more generally
hyperk\"ahler times a flat space).  This is reflected
by the fact that when we talk about holomorphic curves
we have a family of possible choices for that notion, and
therefore the amplitudes vanish due to
higher supersymmetry.  Namely
for every holomorphic curve, we have a $U(1)$
rotation symmetry of the phase of $\zeta$ given by where they intersect
the $S^1$.  We thus get
a family of such curves with no fixed point, and
thus the fermion zero mode along that direction kills
the amplitude\foot{Another way to see this is that when we change
the complex structure, the holomorphic curve seizes to be
holomorphic and we have nothing contributing to the amplitudes.}.
For this reason we modify our geometry so that we have half as much
supersymmetry, i.e. (2,2) supersymmetry in 2 dimensions.

The basic idea is the following:  The BPS particles correspond to holomorphic
M2 branes with respect to different complex structure.   In particular
for a fixed $\theta$, consider a new pair of complex variables given
by
$$w_1=z_1+e^{i\theta} \bar z_2 \qquad w_2=e^{-i\theta}z_2-\bar z_1$$
We can view this as a special case of the twistorial construction
where 
\eqn\twi{w_1=z_1+\zeta \bar z_2\qquad
	w_2=\zeta^{-1} z_2-\bar z_1,}
Consider a new K{\"a}hler form given on ${\bf C}^2$ by
$$-ik_{\theta}=dw_1\wedge d{\overline w}_1+dw_2\wedge d{\overline w}_2$$
Then it is easy to see that
$$-i\, k_\theta= 2\,dz_1\wedge dz_2\cdot e^{-i\theta}-2\, d\bar z_1\wedge d\bar z_2\cdot e^{i\theta}$$
and
$$dw_1\wedge dw_2+d\bar w_1\wedge d\bar w_2=2\,dz_1\wedge dz_2\cdot e^{-i\theta}+2\, d\bar z_1\wedge d\bar z_2\cdot e^{i\theta}.$$
From  the first identity, it is clear that $F(z_1,z_2)=0$ is a Lagrangian subspace
of ${\bf C}^2$ with respect to this K{\"a}hler form.  Moreover, from the second one we see that holomorphic
M2 branes ending on $F=0$, will be such that 
$${\rm Im}(i\, dz_1\wedge dz_2\cdot e^{-i\theta})\big|_{M2}=0$$
$$ 0< k_\theta\big|_{M2}=2\, {\rm Re}(i\, dz_1\wedge dz_2\cdot e^{-i\theta})\big|_{M2}=2\,i\, dz_1\wedge dz_2\cdot e^{-i\theta}\big|_{M2}$$
so that the restriction of $i\,dz_1\wedge dz_2\, e^{-i \theta}\big|_{M2}$
will be real positive. In other words, BPS particles whose central
charge have phase $\theta$ are holomorphic $M2$ branes
with this choice of complex structure.

This construction suggests what we need to do:  We should change
the geometry so that the internal geometry of the brane
is given by an A-brane, with topology of $\Sigma \times S^1$
and where for each $\theta$ the restriction of the K\"ahler
form of the 3-fold to ${\bf C}^2$ gives the structure given above.
In other words the geometry should be fibered in a twistorial
fashion, as discussed above.
In this way the BPS particles will correspond to holomorphic M2 branes which attach
to $S^1$ at specific points given by 
the phase of their central charge.  The exact way this construction
is done would not affect our conclusions, which will
be presented in the next section.    We will discuss
one such construction in section 5.  We find that in this
simplest construction, the $\Sigma \times S^1$ brane though not
quite Lagrangian, behaves as one, since every 2 dimensional subspace of it
(given by $\zeta =const.$) is special Lagrangian.  The
A-model amplitudes are well defined, because
the path-integral is localized on horizontal slices (w.r.t. $\zeta$) for which the Lagrangian condition is satisfied.

\newsec{Derivation Of BPS Degeneracies And Their Jumps}
In this section, using the ideas of the previous section, we show how the topological string captures the BPS degeneracies.  Furthermore
we show an immediate consequence of it is a refined version
of the Kontsevich-Soibelman conjecture \KS .

In the previous section we have shown that a particular twistorial
compactification of $N=2$ theory from 4 to 2 dimensions can yield
a theory with $(2,2)$ supersymmetry.  Namely we found that,
by a chain of dualities, this is equivalent to a theory living
on the D4 brane wrapping a submanifold $\Sigma
\times S^1$ and filling the ${\bf R}^2$.  Topological A-model
string computes F-terms for this $(2,2)$ theory as reviewed in
section 2.  As discussed there, the partition function of the open
topological string receives corrections from holomorphic curves
ending on $\Sigma \times S^1$.  As noted in the previous section the
curve projects to a particular point on $S^1$ and moreover, these are in turn equivalent
to M2 branes ending on the M5 brane.  Thus
the partition function of open topological strings living on the D4
brane has the full information about the quantum
numbers, the degeneracy and the spin of all BPS particles M2 branes ending
on M5 branes.
In particular the corrections to the Chern-Simons theory living
on $\Sigma \times S^1$ is given by
$$Z_{CS}=\langle \prod_{\gamma,j}
O_{j,\gamma}(\theta_\gamma)\rangle $$
where
$$O_{j,\gamma}={\rm exp}\Big[N_{j,\gamma} \sum_n {U_\gamma^{n}\over n} {q^{nj}\over (q^{n/2}-q^{-n/2})}\Big]$$
and $N_{j,\gamma}$ denotes the net number of BPS states
with spin $j$ and electric/magnetic charge given by $\gamma$
and $U_\gamma$ denotes the Wilson loop operator along $\gamma$.
The parameter $q$ is related to the coupling of the topological
string which is in turn the same as the coupling of the CS theory:  $q={\rm exp}(-g_s)$.
Note that these Wilson loops which arise due to holomorphic curves,
are localized on specific location $\theta_\gamma$ on the $S^1$,
due to the fact that BPS states with that given $\gamma$ can be
holomorphic only for that angle as discussed in the previous section.  We thus see that the partition 
function of the topological A-model for this CY
captures the electromagnetic BPS states of the 4d $N=2$ theory
including the information about their spin!  Note that we can
insert arbitrary extra observables (such as by introducing additional branes, etc.)
into this topological theory, and we can
thus in principle extract all the $N_{\gamma,j}$ using topological A-model.

As an application of this construction we are ready to derive the KS conjecture:  For that we view
the Chern-Simons theory on $\Sigma \times S^1$ in the operator
formulation.  Namely we view $S^1$ as time and $\Sigma$ as
space.  Then, as is well known \ref\WittenHF{
  E.~Witten,
  ``Quantum field theory and the Jones polynomial,''
  Commun.\ Math.\ Phys.\  {\bf 121}, 351 (1989).
  %%CITATION = CMPHA,121,351;%%
}\
the classical phase space is given by the space of flat $U(1)$ connections
on $\Sigma$.  This is naturally identified with the Jacobian of $\Sigma$
with the natural symplectic structure giving the Poisson bracket, which
arises from the $AdA$ term in the CS Lagrangian.  If
we parameterize the holonomies of the $U(1)$ along canonical
basis of 1-cycles on $\Sigma$ given by  $(A_i,B_j)$, by the angles $(\eta_i,\phi_j)$,
we have classically
$$\{ \eta_i ,\phi_j \} =\delta_{ij}$$
In the quantum theory this leads to the commutation relation
$$[\eta_i, \phi_j ]=ig_s \delta_{ij}$$
In this basis the holonomy of the operator $U_{\gamma}$, for $\gamma =n^iA_i+m^j B_j$ is given by
$$U_{\gamma}={\rm exp} \big[ i (\eta_i n^i +\phi_j m^j)\big]$$
Moreover note that in the operator formulation we need to time order
the fields.  Since the time is identified with the phase of the BPS states,
this simply means $\theta$ ordering of the operators $O_{j,\gamma}$.
Note that since $[U_{n\gamma},U_{m\gamma}]=0$ there is no
ambiguity when both of these operators appear at the same time $\theta$.
We thus find that the time/$\theta$ ordered operator
$$T(\prod_{\gamma,j}
O_{j,\gamma}(\theta_\gamma))$$
is the evolution operator.   Now consider changing the parameters
of the $N=2$ theory in 4 dimensions.  The BPS states
may get aligned, and cross walls of marginal stability.
If this happens the operator ordering of their contributions
to the Chern-Simons theory gets changed.  However the path-integral
of the A-model, and thus the evolution operator given above is continuous!  Thus the contributions of the BPS
states to the CS theory Wilson loops just after crossing the wall of marginal stability should be the same as the one just before crossing.   As we cross the wall, some phases of the central charges of some BPS states 
get aligned.  Let us focus on two such charges, given by
the classes $\gamma, \gamma' \in H_1(\Sigma, {\bf Z})$.  This means that for the same $\theta $
there are contributions to the topological strings involving $U_\gamma$ and $U_{\gamma'}$.  
If $\langle \gamma ,\gamma'\rangle \not=0$ then the Wilson loops pass through the same point(s)
in $\Sigma$.  This means that
 as a cycle in $\Sigma \times S^1$ the linking of these Wilson loop observables
will change as we cross the wall.   Since the BPS degeneracies
dictate which combinations of them we obtain in the A-model
path-integral (according to
dilog and its generalization) we need to change these degeneracis, in order
for the combinations of CS link amplitudes not to change. 

By inserting extra observables in the CS theory, we can localize
this statement along the twistor circle.  The $g_s\rightarrow 0$ limit of this statement (i.e. the disc amplitudes of the A-model)
leads to a Poisson action on the phase space which reproduces the symplectic
morphism of the KS.   Here we have uncovered the reason for
the appearance of dilogs in their construction.  Moreover, we have a refinement of their statement; namely
we can consider this continuity of the operator for arbitrary $g_s$.  This
captures the spin dependence of the BPS states and leads to a refinement of KS  proposed in 
\guktudor  .  

\newsec{Construction of the A-brane}
Consider the geometry we ended up in section 3, namely
the complex space given by $M={\bf C}^2\times {\bf C}^*$
where ${\bf C}^2$ is parameterized by $w_1,w_2$ and
${\bf C}^*$ by $\zeta$.   These are related to the $z_i$
by
$$w_1=z_1+\zeta \bar z_2\qquad
	w_2=\zeta^{-1} z_2-\bar z_1$$
In this geometry we consider the 3-dimensional submanifold
given by $K=\Sigma \times S^1$ defined by
$$F(z_1,z_2)=0,\quad |\zeta|=1.$$
Let $C\subset {\bf C}^2\times {\bf C}^*$ be a holomorphic curve with boundary in $K$. The projection of
$C$ on the second factor, ${\bf C}^*$, is necessarily a point on the unit circle.\foot{This
is because the $S^1$ is a non-contractible cycle in the geometry.}   All such curves $C$ will end on the $F=0$ curve which
is special Lagrangian inside ${\bf C}^2$, and each one will
map to a point on the $\zeta =e^{i \theta}$ cylinder, with
$\theta$ corresponding to the phase of the corresponding BPS states.

One should ask if the $K$ we
just constructed is Lagrangian, as is generically required
for A-branes.  In order to do this we have
to pick a K\"ahler form.  We equip ${M}$ with the natural flat K\"ahler form
$$	-i\, \omega_{M}=dw_1\wedge d\bar w_1+dw_2\wedge d\bar w_2+ {d\zeta\over\zeta}\wedge {d\bar\zeta\over\bar\zeta},$$
as well as the holomorphic $3$-form
$$\Omega_3={d\zeta\over \zeta}\wedge dw_1\wedge dw_2,$$
making $M$ into a (flat) Calabi-Yau $3$--fold.

In ${M}$, we consider the real codimension $1$ submanifold
$${R}=\{\,\zeta =e^{i\theta},\ \ 0\leq \theta <2\pi\},$$
that is the twistor space ${M}$ restricted to the {equator of the twistor sphere.}
We write $\varpi$ for the symplectic $2$-form
$-i\, \omega_{M}\big|_{R}$.  We notice the following identity
\eqn\give{\varpi= 2\, dz_1\wedge d(e^{-i\theta}z_2)-2\, d\bar z_1\wedge d(e^{i\theta}\bar z_2).}
This identity implies that for fixed $\theta$, 
$$\varpi_\theta\Big|_K =0,$$
as we had noted in section 3.  However we now see that
$\omega_{M}\Big|_K \not=0$, due to the
$z_2dz_1 d\theta e^{-i\theta}$ term, if we consider the non-horizonatal directions,
i.e., directions for which $\theta \not=const$.   In other words,  if there were configurations
in the bulk of the closed worldsheet which had non-constant $\zeta$,
then this would have led to a problem in defining the string amplitudes
in the presence of $K$.
However since all the holomorphic maps are automatically horizontal,
and $K$ is special Lagrangian in those directions,
we have no difficulty in defining the path-integral measure for the
A-model.\foot{The degenerate holomorphic maps also contribute
to the A-model \wittencs, and lead to tree level Chern-Simons 
Lagrangian.  These configurations will not be horizontal; however
they will not affect our argument because the pair of boundaries
ending on $K$ in that case will cancel, as they correspond
to open strings of infinitesimal width.}  One way to understand
this is to consider rescaling the K\"ahler metric in the $\zeta$
direction by a large number $t$:
$$\omega_t = i\, t\,{{d\zeta}\over{\zeta}}\wedge {{d\bar \zeta}\over{\bar \zeta}}+ i\, dw_1(\zeta)\wedge d\bar w_1(\bar\zeta)+i\, dw_2(\zeta)\wedge d\bar w_2(\bar\zeta),$$
and evaluate the path--integral for the corresponding $\sigma$--model, with Dirichlet boundary conditions on $K$, in the limit $t\rightarrow\infty$. 

We perform first the integral over the non-zero modes of the field $\zeta$, for a fixed configuration of the fields $z_1$, $z_2$,
and boundary condition on $\zeta$ given by $\big|\zeta\big|_{\partial D}\big|=1$.
The integral is Gaussian, and the result is an effective action for the fields $z_1$, $z_2$ of the form
$$	\partial^\mu(z_1+e^{i\theta}\, \bar z_2)\partial_\mu(\bar z_1+ e^{-i\theta}\, z_2)+	\partial^\mu(e^{-i\theta}\,z_2- \bar z_1) \partial_\mu(e^{i\theta}\,\bar z_2-  z_1)+ O\left({1\over t}\right)$$
where $\zeta= e^{i\theta}$ is the zero mode of $\zeta$, and the term $O(1/t)$ comes from the integral over the non-zero modes. Notice that all the BRST non-invariance of this result is in the $O(1/t)$ term.
  Dropping the $O(1/t)$ terms, we end up with a  path-integral in $w_1$, $w_2$ which is the standard one for the $A$-model in flat space, with now a boundary condition which is a standard special Lagrangian brane in one less dimension, namely
 the curve $F(z_1,z_2)=0$ in $\bf{C}^2$.
The path integral in $z_1$, $z_2$ is then well-defined as a topological amplitude. We get a non-trivial contribution only for the discrete set of $\theta$'s for which there are BPS states with that central charge phase. Then the final integration over the $\zeta$ zero-modes, $\theta$, may be replaced by a discrete sum.
  \smallskip
  
In this way we see that, at least for large $t$, the BPS-counting path integral should be well-defined as a topological amplitude. As we have argued, it corresponds to a specific correlator in the CS theory on $\Sigma\times S^1$. The fact that the length of the $S^1$ is taken to $\infty$ is
unimportant since CS is a strictly topological theory insensitive to metric deformations.
This concludes the argument that this brane leads to
sensible path-integral contributions for the A-model.  It would be
important to make this argument fully rigorous.

Even though we have a construction which captures
the essence of what we need, it would have been more
satisfactory if we had found
a $K$ which is special Lagrangian.  It is not difficult to see that
modifying the K\"ahler metric in the particular complex geoemtry we introduced will not lead to a Lagrangian
 $K$.
It would be important to see if a relaxation of some of the other
assumptions may lead to one.  In fact the most natural
choice for the restriction of the K\"ahler form to $|\zeta|=1$, which 
has the required property as well as vanishing on $K$ is
$e^{-i\theta}dz_1\wedge dz_2+c.c.$.  This form is manifestly
not closed.   So we can make $K$ Lagrangian at the expense
of making the geometry non-K\"ahlerian.  This suggests that if we wish to have the 2-form
vanish on $K$, we have to allow
generalized
complex and K\"ahler structures
\ref\HullVW{
  C.~M.~Hull, U.~Lindstrom, M.~Rocek, R.~von Unge and M.~Zabzine,
  ``Generalized Kahler geometry and gerbes,''
  arXiv:0811.3615 [hep-th].
  %%CITATION = ARXIV:0811.3615;%%
}\ref\KapustinGV{
  A.~Kapustin and Y.~Li,
  ``Topological sigma-models with H-flux and twisted generalized complex
  manifolds,''
  arXiv:hep-th/0407249.
  %%CITATION = HEP-TH/0407249;%%
}\ref\PestunRJ{
  V.~Pestun,
  ``Topological strings in generalized complex space,''
  Adv.\ Theor.\ Math.\ Phys.\  {\bf 11}, 399 (2007)
  [arXiv:hep-th/0603145].
  %%CITATION = 00203,11,399;%%
},
which is natural if we have various kinds of fluxes turned on
in the bulk.     In fact a
construction similar to what we need is already
done in \ref\TomasielloEQ{
  A.~Tomasiello,
  ``New string vacua from twistor spaces,''
  Phys.\ Rev.\  D {\bf 78}, 046007 (2008)
  [arXiv:0712.1396 [hep-th]].
  %%CITATION = PHRVA,D78,046007;%%
}\ (see also \ref\PopovNX{
  A.~D.~Popov,
  ``Hermitian-Yang-Mills equations and pseudo-holomorphic bundles on nearly
  %Kaehler and nearly Calabi-Yau twistor 6-manifolds,''
  arXiv:0907.0106 [hep-th].
  %%CITATION = ARXIV:0907.0106;%%
}) where it is shown how to construct
a supersymmetric geometry using the twistor space with
an $SU(3)$ structure with a non-closed analog of K\"ahler form.  The $SU(3)$ structure uses
a non-conventional complex structure on the
twistor space which is not integrable and satisfies
equations very much like what we need for our construction.
Indeed this non-conventional complex structure for
twistor space was already introduced in \ref\salamon{J. Eells and S. Salamon, ``Twistorial construction of harmonic maps of surfaces into
four-manifolds," Ann. Scuola Norm. Sup. Pisa Cl. Sci. {\bf 4}
no. 4, 12 (1985) 589.}\
in order to relate the propblem of constructing harmonic maps from surfaces
into four manifolds to holomorphic maps in the associated twistor space!
It seems reasonable to expect that the most natural construction
for our problem involves such a choice.
We leave this for future studies. 

\newsec{Conclusion}

We have seen that the BPS states of $N=2$ gauge theories in 4 dimensions are captured
by open A-model topological string amplitudes.  On the face of it, this seems in line
with other appearances of BPS counting in topological strings \govai . However
in those cases one typically computes only the electric BPS degeneracies.  Here by a series
of dualities we have been able to also realize counting of arbitrary electric
and magnetic states in an associated A-model with higher bulk supersymmetry,
whose supersymmetry is reduced by the presence of a Lagrangian brane.
Thus for example for the A-model on toric geometry, the degeneracy of D2 and D4 branes
get mapped to open A-model amplitudes realized on a twistorial space where the mirror of the toric geometry defines a brane. 

We have offered a twistorial construction which captures
what we need for our application.  However, as we have noted,
the condition of the A-brane being Lagrangian is not easy to satisfy in
a Calabi-Yau background.  It seems a
better construction may require going to non-K\"ahlerian geometries.

We have mainly focused on the case where the
Seiberg-Witten curve is embedded in locally flat space.  However
there are more general cases (such as branes wrapping $T^*C$) which are not
of this type.  It should be possible to extend the twistorial constructions
to these cases as well \ref\future{S. Cecotti, A. Neitzke and C. Vafa, work in progress}.

As we have noted in the context of A-model topological string it
is natural to consider an even more refined invariant, having to do
with taking the multiplicty of M5 branes  to be $K> 1$.  The physical
interpretation of the monodromy in this case would be interesting
to explore.  In particular the Wilson loop holonomies will be
elements of $U(K)$.

It would be interesting to connect this work, with the approach taken in 
\GMN\ in deriving the KS conjecture.  In particular it would be interesting
to translate  the data for corrections to hyperk\"ahler metric in the setup
we have considered here.

This project is an offshoot of our work \future ,
in attempting to understand the meaning of KS monodromy.  In analogy
with the 2d case \cecva , one expects that the monodromy and its eigenvalues
encode R-charge of chiral operators at conformal fixed points.  
The fact that  monodromy captures R-charges in 2-dimensional theories was the
highlight of those results.  It is natural to expect to have a similar story for 4d $N=2$ SCFT's.  
There are already encouraging signs in that direction \future .

\vglue 2cm

We would like to thank Andy Neitzke for many helpful conversations and for
an ongoing project on a related topic.  We would also like to thank Martin Rocek and Alessandro Tomasiello
for helpful discussions on twistor spaces.
SC would like to thank the High Energy Theory group at Harvard for its hospitality.
CV would  like to thank the Simons
Center for Geometry and Physics for hospitality during the
7th Simons workshop on Physics and Mathematics.

The research of CV was supported in part by NSF grant PHY-0244821.

\listrefs

\end